\documentclass[aps,pra,amsmath, amsfonts, amssymb, superscriptaddress, twocolumn, showpacs]{revtex4}

\usepackage{graphicx}
\usepackage{setspace}
\usepackage{wasysym}
\usepackage{color}
\usepackage{nicefrac}

\def\a{s}
\def\b{s}
\renewcommand{\vec}[1]{\mathbf{#1}} 
\newcommand{\add}[1]{\if\a\b{{\color{red} #1}}\else{#1}\fi}
\newcommand{\comm}[1]{\if\a\b{{\color{blue} #1}}\else{#1}\fi}

\newcommand{\mean}{\operatorname{mean}}
\newcommand{\scat}{\sigma_\text{tot}}
\newcommand{\wop}{\omega_\text{op}}

\newcommand{\lop}{\lambda_\text{op}}

\newcommand{\citeasnoun}[1]{Ref.~\onlinecite{#1}}

\renewcommand{\eqref}[1]{Eq.~(\ref{#1})}

\newcommand{\secref}[1]{Sec.~\ref{sec:#1}}

\newcommand{\figref}[1]{Fig.~\ref{fig:#1}}

\renewcommand{\Re}{\operatorname{Re}}
\renewcommand{\Im}{\operatorname{Im}}
\newcommand{\J}{\mathcal{J}}

\begin{document}
\def\linefigwidth{0.5\textwidth}
\def\smalllinefigwidth{0.35\textwidth}
\def\smallerlinefigwidth{0.25\textwidth}
\def\largelinefigwidth{0.5\textwidth}

\title{A diameter--bandwidth product limitation of isolated-object cloaking}

\author{Hila Hashemi}
\affiliation{Department of Mathematics, Massachusetts Institute of Technology, Cambridge, MA 02139}
\author{Cheng Wei Qiu}
\affiliation{Department of Electrical and Computer Engineering, National University of Singapore, Singapore 117576, Singapore}
\author{Alexander P. McCauley}
\affiliation{Department of Physics, Massachusetts Institute of Technology, Cambridge, MA 02139}
\affiliation{WiTricity Corporation, Watertown, MA 02472}
\author{J.~D.~Joannopoulos}
\affiliation{Department of Physics, Massachusetts Institute of Technology, Cambridge, MA 02139}
\author{Steven~G.~Johnson}
\affiliation{Department of Mathematics, Massachusetts Institute of Technology,
 Cambridge, MA 02139}

\begin{abstract}
We show that cloaking of isolated objects is subject to a diameter--bandwidth product limitation: as the size of the object increases, the bandwidth of good (small cross-section) cloaking decreases inversely with the diameter, as a consequence of causality constraints even for perfect fabrication and materials with negligible absorption.  This generalizes a previous
result that \emph{perfect} cloaking of isolated objects over a nonzero bandwidth violates causality.  Furthermore, we demonstrate broader causality-based scaling limitations on any bandwidth-averaged cloaking cross-section, using complex analysis and the optical theorem to transform the frequency-averaged problem into a single scattering problem with transformed materials.
\end{abstract}

\maketitle 
\section{Introduction}
In this work, we extend the result that perfect isolated-object cloaking is impossible over a nonzero bandwidth~\cite{Pendry06, Miller06} to show that even \emph{imperfect} cloaking of isolated objects necessarily has a bandwidth that decreases with the size of the object. More generally, we show that the cloaking efficiency (the ratio of total cross section to geometric cross section) must worsen proportional to the diameter when averaged over any finite bandwidth for cloaking of isolated objects in air (or any medium of negligible loss and dispersion). Unlike our previous proof that sensitivity to imperfections worsens with diameter at individual frequencies~\cite{Hashemi11}, the result in this paper holds even for perfect fabrication and for materials with negligible absorption over the desired bandwidth.  Our proof involves some unusual mathematical techniques. First, we equate the frequency-averaged problem to a single scattering problem at a complex frequency with the help of a version of the optical theorem. Second, we map the complex-frequency problem to an equivalent real-frequency problem with transformed materials. We can then use the fact that $\Im \varepsilon$ and $\Im \mu$ are $> 0$ for any physical causal material at $\Re \omega$ and $\Im \omega > 0$~\cite{Landau84:electro} to analyze an ``effective'' absorption loss in a manner similar to our previous work~\cite{Hashemi11}.  Finally, in \secref{num} we numerically verify this scaling in an example of a spherical cloak that is a perfect Pendry cloak at one frequency but has causal dispersion.

  The idea of transformation-based invisibility cloaks was proposed in 2006 \cite{Pendry06}, a fascinating idea that was followed by many theoretical works~\cite{Leonhardt06NJP, Schurig06OE, Cummer06, Qiu09, Chen07PRB, Kwon08,Jiang08, Kante08, Ma09, Yan07, Ruan07,Huang07, Zhang08APL, Cai07NP, Zolla07,Tucker05, Nicolet08, Chen08JAP, Leonhardt09, Baile09, Argyropoulos10, Baile10OE, Han10, Han11} and several experimental demonstrations of cloaking of isolated objects at one frequency~\cite{Schurig06, Smolyaninov08, Kante09, Liu09APL}. However, as pointed out by Pendry using a speed-of-light argument~\cite{Pendry06} and by Miller with a more formal approach~\cite{Miller06}, \emph{perfect} cloaking of an isolated object in vacuum over a non-zero bandwidth is impossible due to causality. This severe limitation helped inspire the idea of ground-plane cloaking \cite{Li08,ErginSt10OE, Xu09, Baile10, Landy10}  and its experimental demonstrations~\cite{Liu09, Ma09OE, Gabrielli09, Valentine09,Lee09, ErginSt10, Ma10, Zhang11, Chen11, Gharghi11}, which circumvents such causality limitations (although it is still subject to other practical scaling difficulties~\cite{Hashemi10,Hashemi11}). However, the Pendry and Miller proofs say little about imperfect cloaking (a nonzero but small scattering cross-section).  By continuity, if near-perfect cloaking is attained at a single frequency, ``good'' cloaking (total cross-section bounded by some given fraction of the geometric cross-section) must persist over some finite bandwidth. We show that causality imposes an even stronger constraint than forbidding perfect cloaking over a finite bandwidth: for a bandwidth-limited cloak, we show that causality constraints imply that the bandwidth of good cloaking scales inversely with the object diameter.  (The impossibility of perfect cloaking over a finite bandwidth follows from our results as a special case.) Moreover, our proof holds for isolated objects in any transparent medium (negligible loss and dispersion), not just in vacuum.  

The key to our proof in \secref{proof} is the transformation of a frequency-averaged scattering problem (weighted by a Lorentzian window for convenience), which is hard to analyze, to a \emph{single} complex-frequency scattering problem that is easy to analyze by relating the complex frequency to equivalent complex materials. In order to make this transformation, we rely on the optical theorem~\cite{Jackson98}, which relates the total cross section to the imaginary part of a forward scattering amplitude, since the latter is a causal linear response and hence an analytic function in the upper-half complex-frequency plane~\cite{Landau84:electro}. A review of the optical theorem and its history and applications can be found in \cite{Newton76}. The use of the optical theorem, and similar relationships based on conservation of energy, to apply complex analysis to scattering problems is most common in quantum field theory~\cite{Peskin95}. Similar in spirit to this
paper, the optical theorem is also used in the ``ITEP sum rules'' of
quantum chromodynamics to relate integrals of scattering cross
sections multiplied by Lorentzian windows (or powers thereof) to
scattering amplitudes at single points in momentum space via contour
integration~\cite{Peskin95}.  On another related note, contour
integration of the imaginary part of Green's functions (and other
scattering amplitudes) is also a key technique for computing Casimir
interactions in quantum field theory~\cite{Johnson11}, where the
relationship between the imaginary part of the Green's function and
the field fluctuation statistics (the fluctuation--dissipation
theorem) is, like the optical theorem, derived from
energy-conservation considerations~\cite{Landau:stat}.

\section{Derivation}
\label{sec:proof}

A transformation-based cloak, as depicted in Fig. 1, works by mapping an object (in ``physical'' space $X$) to a single point (in ``virtual space'' $X'$) and the cloak region (volume $V_c$) to empty space (volume $V_c'$), via a coordinate transformation that is the identity outside the cloak. If the ambient materials are $\varepsilon_a$ and $\mu_a$, then an ideal cloak is obtained by constructing the materials $\varepsilon = \J\varepsilon_a\J'/\det \J$ and $\mu = \J\mu_a\J'/\det \J$ in the cloak, where $\J$ is the Jacobian matrix of the transformation. In our previous work~\cite{Hashemi11} we showed that the cloaking problem at one frequency becomes increasingly difficult as the size of the object being cloaked increases. In particular, we proved that the imperfections due to absorption losses and random fabrication disorder must decrease asymptotically with the object diameter in order to maintain ``good'' cloaking performance: for the total (scattering + absorption) cross-section to be less than a given fraction $f$ of a geometric cross section $s_g$. To prove these results, we assumed bounds on the attainable refractive index contrast in the cloak, $b<n/n_a<B$ ($n=\sqrt{\varepsilon\mu}$), which we showed to be equivalent to bounds on the singular values of $\J$. Using these bounds, we were then able to bound the field amplitudes in the cloak in terms of the field amplitudes in the virtual space (which are a constant for incident planewaves); such bounds, in turn, impose bounds on the allowed absorption losses and random disorder. In particular, we showed in the case of absorption losses that $\Delta \Im \varepsilon$  scales proportionally to $\frac{s_g}{V'_c}$, a measure of the inverse diameter.

We now wish to analyze the effect of other imperfections, especially inevitable material dispersion, on the cloaking performance, even in the idealized case of perfect fabrication and negligible absorption. In particular, we suppose that one is interested in the \emph{average} total cross section $\scat$ over a bandwidth $\Delta\omega$ around an operating frequency $\wop$, which (for reasons described below) is convenient to define via a Lorentzian averaging weight as:

\begin{equation}
\langle\scat \rangle_\Delta \omega = \int_{-\infty}^{\infty} \scat\frac{\Delta \omega/\pi}{(\omega-\wop)^2+\Delta \omega^2}d\omega
\label{meanscat}
\end{equation}

As we increase the diameter $d$ of the object and cloak, keeping the materials and transformation mapping fixed and simply rescaling the whole system, we show that the cloaking problem becomes increasingly difficult, in the sense that $\langle \scat \rangle_{\Delta \omega}/s_g \sim d$ (where $\sim$ means proportional to), with $s_g$ being a geometric cross sectional area of the object.  (This scaling breaks down when $\scat / s_g$ is no longer small, i.e. when it is no longer an approximate cloak.) The only assumptions are that the ambient medium $\varepsilon_a$ and $\mu_a$ is approximately lossless and dispersionless over the bandwidth of interest (e.g. for cloaking in air) and that the attainable refractive index contrast (eigenvalues of $\sqrt{\varepsilon \mu / \varepsilon_a \mu_a}$) is $\leq B$ for some finite bound $B$ (similar to our previous work~\cite{Hashemi11}).

Our analysis in the following subsections constitutes the following steps.  First, in \secref{fa}, using complex analysis combined with the optical theorem, we relate the frequency average of \eqref{meanscat} to a \emph{single} scattering problem at a \emph{complex} frequency so that we no longer need to consider the cross section at many frequencies at once.  Second, in order to understand the precise meaning of this complex-frequency scattering problem, in \secref{optical} we derive a variant of the optical theorem that is particularly easy to analyze.  Third, in \secref{freqtomat} we relate this complex-frequency scattering problem to an equivalent scattering problem at a \emph{real} frequency with transformed complex \emph{materials}. Fourth, in \secref{dispersion} we use analysis similar to our proof in~\cite{Hashemi11} to show the ``losses'' introduced by these effective complex materials must scale with diameter, and hence mean $\scat$ must scale with diameter. Finally, for the special case of a bandwidth-limited cloak (a cloak that is very good at one frequency but spoiled at other frequencies by material dispersion), we show that the bandwidth must narrow inversely with diameter in \secref{bandwidth-scaling}.

\begin{figure}[t]
\centering
\includegraphics[width=1.0\columnwidth]{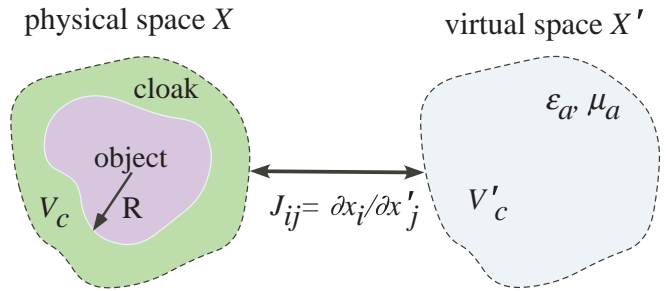}
\caption{Schematic of transformation-based isolated-object cloak, which works by mapping the object and cloak in physical space $X$ to a virtual space $X'$ in which the object is a single point. The transformation laws of Maxwell's equations~\cite{Ward96, Pendry06} then dictate the required $\varepsilon$ and $\mu$ materials in the cloak volume to produce equivalent solutions in $X$ and $X'$.}
\label{fig:cloak-iso}
\end{figure}

\subsection{Frequency average of the scattering cross-section}
\label{sec:fa}

By the optical theorem~\cite{Jackson98}, $\scat(\omega) = \Im f(\omega)$ where $f$ is a forward-scattering amplitude (at least for the case of an incident planewave, but a generalization is given in the next section for any incident field), and so
\begin{equation}
\langle \scat \rangle_{\Delta \omega}= \Im \int_{-\infty}^{\infty} f(\omega) \frac{\Delta \omega/\pi}{(\omega-\wop)^2+\Delta \omega^2}d\omega.
\end{equation}
Because $f$ is a causal linear response, it is analytic for $\Im \omega>0$, and therefore we can perform a contour integration to obtain $\langle \scat \rangle_{\Delta \omega}$ in terms of the residue of the single pole of the Lorentzian in the upper-half complex plane:
\begin{equation}
\langle \scat \rangle_{\Delta \omega} = \Im f(\wop + i\Delta\omega) .
\end{equation}
This corresponds to a single scattering problem at a complex frequency, analyzed in more detail below.

\subsection{The optical theorem and analytic continuation to complex $\omega$}
\label{sec:optical}
Consider any finite-volume scatterer (in a linear time-invariant system), described by some change $\Delta\varepsilon$ and $\Delta\mu$ in the permittivity and the permeability compared to the ambient medium.  Let the incident field (in the absence of the scatterer) be described by a six-component vector field $\psi_{\text{inc}}= \left( \begin{array}{cr} \vec E_{\text{inc}} \\ \vec H_{\text{inc}} \end{array} \right)$.  In the presence of the scatterer, this is modified to a new total field $\psi = \psi_{\text{inc}}+\psi_{\text{scat}}$, the sum of the incident and scattered fields.  The charge perturbations in the scatterer are described by bound electric and magnetic polarization currents $\Delta\vec J = -i\omega \Delta\varepsilon \vec E$ and $\Delta\vec K = -i\omega \Delta \mu \vec H$, respectively, which can be combined into a six-component current field $\xi = \left(\begin{array}{c}\Delta\vec J\\ \Delta\vec K\end{array}\right)$.  Abstractly, we can write $\xi = A \psi_\text{inc}$ for some linear operator $A$ relating the incident fields to the induced currents, and causality (currents come after fields) implies that $A$ is an analytic function in the upper-half complex-$\omega$ plane~\cite{Landau84:electro}.

Physically, the scattered field is the field produced by these
oscillating induced currents $\xi$ in the scatterer, and the
interactions of the currents and fields provide a simple way to
characterize the absorbed and scattered powers. The total absorbed
power $P_{\text{abs}}$, assuming an ambient medium with negligible
dissipation, equals the total time-average ($\psi$) incoming Poynting
flux:
\begin{equation}
\begin{split}
P_\text{abs} &= \frac{1}{2} \Re \int (\vec E^* \cdot \Delta\vec J + \vec H^* \cdot \Delta\vec K) \\
             & = \frac{1}{2} \Re \left<\psi,\xi\right> \\
             & =\frac{1}{2}\Re\left<\psi_{\text{inc}},\xi\right>+\frac{1}{2}\Re\left<\psi_{\text{scat}},\xi\right>,
\end{split}
\end{equation}
where we have defined the inner product $\left<\cdots,\cdots\right>$.
Similarly, the scattered power is the work done by the currents on the scattered field: $-\frac{1}{2}\Re \left<\psi_{\text{scat}},\xi\right>$. Therefore, the total ($\mathrm{absorbed} + \mathrm{scattered}$) power is:
\begin{equation}
\begin{split}
P_{\text{tot}}  & = P_{\text{abs}}+P_{\text{scat}} \\
                & = \left(\frac{1}{2}\Re\left<\psi_{\text{inc}},\xi\right>+\frac{1}{2}\Re\left<\psi_{\text{scat}},\xi\right>\right) -\frac{1}{2}\Re \left<\psi_{\text{scat}},\xi\right> \\
                & = \frac{1}{2}\Re \left <\psi_{\text{inc}},\xi\right> \\
                & = \frac{1}{2} \Re \left<\psi_{\text{inc}}, A\psi_{\text{inc}}\right> \\
                &  = \Im f(\omega),
\end{split}
\end{equation}
where we have defined $f(\omega) =  \frac{i}{2} \left<\psi_{\text{inc}},A\psi_{\text{inc}}\right>$.
 The interpretation of $f$ as a forward-scattering amplitude when $\psi_\text{inc}$ is a planewave is actually irrelevant to our proof, but it follows from the fact that $f$ in that case is simply the forward-planewave Fourier component of the field $\sim i\xi/\omega$ corresponding to the currents $\xi$.

The key question, for our purposes, is to understand what $f$ looks like at a complex frequency.  Suppose we have homogeneous, lossless ambient medium with an incident planewave $\psi_\text{inc} = \psi_0 e^{i\omega x/c}$ propagating in the $+x$ direction for a constant amplitude $\psi_0$.  Then
\begin{equation}
f(\omega) = \frac{i}{2} \left<\psi_0, e^{-i\omega x/c} A(\omega) e^{+i\omega x /c} \psi_0 \right>,
\end{equation}
moving all the $x$ dependence to the right-hand side of the inner product.  At a complex frequency $\omega \to \omega + i\gamma$ in the upper-half plane ($\gamma > 0$), $A(\omega+i\gamma)$ is analytic by causality, while the planewave terms are analytic everywhere (and become exponentially decaying or growing waves in space at complex $\omega$). So, $f$ at a complex frequency represents an overlap with an exponentially growing field of the currents produced (via $A$) by an exponentially decaying source.  In the next section, we clarify this picture further by relating the scattering operator $A$ at a complex frequency to scattering at a real frequency with complex materials.

In order to perform contour integrations of $f(\omega)$ multiplied by a Lorentzian, it is not enough for $f$ to be analytic, however: it must also be bounded as $|\omega|\to\infty$ so that we can close the contour above.  This fact is used extensively in quantum field theory~\cite{Peskin95}. Intuitively, this occurs because the the exponential growth of $e^{+\gamma x/c}$ cancels the exponential decay of $e^{-\gamma x/c}$.  Mathematically, as $|\omega|\to\infty$ the $\Delta\varepsilon$ and $\Delta\mu$ must vanish (since all susceptibilities vanish in the limit of infinite frequency~\cite{Landau84:electro, Jackson98}), and so $A$ simplifies: to lowest order for weak scatterers (i.e., in the first Born approximation), $\xi \approx A\psi_\text{inc} \approx  \left( \begin{array}{cr} -i\omega \Delta\varepsilon \vec E_{\text{inc}} \\ -i\omega \Delta\mu \vec H_{\text{inc}} \end{array} \right)$, in which case the exponential factors exactly cancel.   We will use a similar procedure, below, to analyze the effects of small imperfections in the cloak due to material dispersion.

\subsection{From complex frequencies to complex materials}
\label{sec:freqtomat}
Combining the previous two sections, we can now relate the frequency-averaged scattering cross-section, for an incident planewave (say in the $x$ direction to the solution of a single scattering problem at a single complex frequency:
\begin{equation}
\langle \scat \rangle_{\Delta \omega} = \frac{1}{2} \Re \left<\psi_0, e^{-i\omega x/c} A(\omega) e^{+i\omega x /c} \psi_0 \right>,
\label{transformed-scat}
\end{equation}
evaluated at $\omega = \wop + i\Delta\omega$.  The $A(\omega)
e^{+i\omega x /c} \psi_0$ represents the induced currents $\xi$, for
the materials evaluated at the complex $\omega$, in response to an
exponentially decaying incident planewave (multiplied by a constant
amplitude $\psi_0$).  The interpretation of this problem is simplified
by the fact that a complex frequency is mathematically equivalent to
using modified \emph{materials} at a real frequency $\wop$. In particular,
consider Maxwell's equations in the frequency domain:
\begin{align}
\nabla \times \vec E & = -i\omega\mu\vec H - \vec K ,  \\
\nabla \times \vec H & = + i \omega \varepsilon \vec E + \vec J ,
\end{align}
where $\vec J$ and $\vec K$ are (free) electric and magnetic current
densities, respectively. For our complex $\omega$, the $i\omega\varepsilon\vec E$
term becomes $i (\wop+ i\Delta
\omega) \varepsilon(\wop+i\Delta \omega) \vec E = i
\left[\varepsilon(\wop+i\Delta\omega).(1+i\frac{\Delta
    \omega}{\wop})\right] \wop \vec E$, which looks exactly like the term for
a real frequency $\wop$ at a modified permittivity $\tilde\varepsilon$
\begin{equation}
\tilde{\varepsilon}(\wop) =
\varepsilon(\wop+i\Delta\omega).\left(1+i\frac{\Delta \omega}{\wop}\right).
\end{equation}
Similarly for $\mu \to \tilde{\mu}$.

Thus, operating at a complex frequency is equivalent to \emph{two}
changes in the materials. First, the multiplication by
$1+i\Delta\omega/\wop$ corresponds to an effective absorption added
throughout all space (including the ambient medium).  Second we
evaluate $\varepsilon$ and $\mu$ at $\wop+i\Delta\omega$ rather than at
the real frequency $\wop$, which will change their values in the
presence of material dispersion.

\subsection{Consequences of dispersion on frequency-averaged scattering}
\label{sec:dispersion}
So far, we have shown that the problem of finding the frequency-averaged scattering cross section is equivalent to solving a certain scattering problem at a single complex frequency, which in turn is equivalent to a scattering problem at a single \emph{real} frequency $\wop$ with modified complex materials $\tilde\varepsilon$ and $\tilde\mu$. In this section, we analyze the consequences of those effective material modifications, divided into two separate material changes as discussed above, for cloaking performance.

\subsubsection{Irrelevance of artificial absorption}
\label{sec:false}
The first change in the materials is the multiplication of $\varepsilon$ and $\mu$ by $1+ i\frac{\Delta\omega}{\wop}$. But this change does not
hurt the cloaking performance because it is done uniformly \emph{everywhere} in space.  More explicitly, if $\varepsilon = \J \varepsilon_a \J^T/\det
\J$ is a valid transformation-based cloak, so is $\varepsilon.(1+\frac{\Delta
\omega}{\wop}) = \J \varepsilon_a \dot (1+\frac{\Delta\omega}{\wop})\J^T/\det\J$,
except that this is now a transformation-based cloak for a lossy ambient medium
$\varepsilon_a. (1+i\frac{\Delta\omega}{\wop})$, and similarly for $
\mu$.  Therefore this change in the materials leaves the scattering cross-section invariant.

\subsubsection{Impact of material dispersion}
\label{sec:real}

The second change in the materials is that we need to evaluate $\varepsilon$ 
and $\mu$ at the complex frequency $\wop+i\Delta\omega$ instead of $\wop$, and here the presence of material dispersion
(unavoidable for any material other than vacuum) acts to spoil the
cloak.  In particular, as reviewed in the Appendix, causality and other fundamental principles imply that $\Im \varepsilon$ and $\Im
\mu$ are both strictly \emph{positive} at $\wop + i\Delta\omega$ for $\Delta\omega>0$~\cite{Landau84:electro}, corresponding to an unavoidable
additional absorption in the complex-frequency scattering problem.  Unlike the
artificial absorption in the previous section, this is an absorption
\emph{defect} introduced only in the cloak: $\varepsilon$ (both $\Im$ and $\Re$) differs from the cloaking transformation of the ambient medium by
$$
\Delta\varepsilon = \left[ \varepsilon(\wop + i\Delta\omega) - \varepsilon(\wop)\right] \left(1+ i\frac{\Delta\omega}{\wop}\right),
$$
and similarly for $\mu$. The key assumptions here are that the ambient medium has no dispersion over the given bandwidth, so that the ideal cloaking transformation at the complex $\omega$ is given by the second term in $\Delta\varepsilon$, and that the ambient medium is lossless, so that $\varepsilon(\wop)$ is real and $\Im\Delta\varepsilon > 0$.  More generally, it is sufficient for the dispersion and loss of the ambient medium to be small enough, compared to the dispersion of the cloak materials, such that $\Im\Delta\varepsilon$ is $>0$.

We can analyze the consequences of this effective absorption imperfection similarly to our previous work~\cite{Hashemi11}
It is convenient to first transform the imperfections to virtual space (as shown in \figref{cloak-iso}) where the scattering problem is easier, because in the absence of $\Delta\varepsilon$ and $\Delta\mu$ we have a planewave in a homogeneous medium.  In particular, given an upper bound $B$ on the index contrast as discussed above, one obtains $\Delta \varepsilon' \geq \Delta \varepsilon / B$ in virtual space, and similarly for $\mu$~\cite{Hashemi11}.  Furthermore we are only interested in the regime in which we have a good cloak---once the imperfections become so large as to make the cloak useless, all of the scaling relations break down and the problem is no longer interesting.  This is the regime in which $\Delta\varepsilon$ and $\Delta\mu$ are small, and therefore virtual space at complex $\omega$ is equivalent to a homogeneous medium with small imperfections and perturbative methods are applicable. In particular, the lowest-order scattering current (in virtual space) is simply $\vec J' = -i\wop \Delta \varepsilon' \vec E_\text{inc}$, and similarly for $\vec K'$.  In the notation of \secref{optical}, $\xi' = A'\psi_\text{inc} \approx  \left( \begin{array}{cr} -i\omega \Delta\varepsilon' \vec E_{\text{inc}} \\ -i\omega \Delta\mu' \vec H_{\text{inc}} \end{array} \right)$.  Substituting this into \eqref{transformed-scat}, we find that the exponential factors $e^{\pm \Delta\omega x/c}$ exponential factors exactly cancel, leaving:
\begin{equation}
\langle \scat \rangle_{\Delta \omega} \approx \frac{1}{2} \int_{V_c'} \left[ |\vec E_0|^2 \Im \Delta\varepsilon'
+ |\vec H_0|^2 \Im \Delta\mu' \right] .
\end{equation}

This has two main consequences.  First, $\mean\scat > 0$ for $\Delta\omega>0$ since $\Im \Delta\varepsilon$ and $ \Im \Delta\mu$ are strictly positive as noted above: even if the cloak is a perfect cloak at $\wop$, it is imperfect when averaged over any non-zero bandwidth. This is, therefore, an alternative proof of the results of Pendry~\cite{Pendry06} and Miller~\cite{Miller06} that cloaking of isolated objects over a non-zero
bandwidth is impossible for physical, causal materials.  Second, exactly as we showed for other imperfections in previous work~\cite{Hashemi11}, it immediately follows that $\langle \scat \rangle_{\Delta \omega}$ grows $\sim V_c' \sim V_c \sim V_o$ and hence the frequency-averaged cloaking efficiency $\langle \scat \rangle_{\Delta \omega} / s_g$ scales proportionally to a mean diameter $V_o / s_g$.  This is the central result of our proof, but
to better understand its consequences we consider some special cases
in the next section.

\subsection{Scaling of cloaking bandwidth with diameter}
\label{sec:bandwidth-scaling}

As shown in the previous section, the fractional cross-section $\mean
\scat / s_g$, averaged over a bandwidth $\Delta\omega$ around $\wop$, must scale
proportional to the diameter $d$, at least as long as $\langle \scat \rangle_{\Delta \omega} /
s_g \ll 1$ (i.e. until cloaking breaks down completely).  There are
two possible sources of this linear scaling, depending on whether
$\langle \scat \rangle_{\Delta \omega}$ is limited by the cross-section at $\wop$ (due to imperfections in the cloak at the design frequency) or by the bandwidth of a dip in the cross-section around
$\wop$ (due to material dispersion degrading a near-perfect single-frequency cloak).  In the former case, the physical mechanism is simply that the
losses due to imperfections at $\wop$ scale with diameter, as we already proved in \citeasnoun{Hashemi11}.  In the latter case, however, it leads to a new prediction:
in a bandwidth-limited cloak, the bandwidth must narrow as the object
diameter increases.  In fact, we show in this section that the bandwidth generically narrows inversely with the diameter in this case.

First, let us consider the bandwidth scaling from generic dimensional
considerations.  Because of material dispersion, one expects good
cloaking to only be possible in some limited bandwidth $\sim \Gamma$
around some design frequency $\wop$, in which case a small $\mean
\scat/s_g$ requires $\Delta\omega \ll \Gamma$.  If we Taylor-expand
$\langle \scat \rangle_{\Delta \omega}/s_g$ in $\Delta\omega / \Gamma \ll 1$, we would generically expect an expansion of the form:
$$
\langle \scat \rangle_{\Delta \omega}/s_g = \scat(\wop)/s_g + C \Delta\omega/\Gamma + O[(\Delta\omega/\Gamma)^2]
$$ for some coefficient $C$, where $C$ is generically $>0$ if $\wop$ is chosen to be a minimum of $\scat$. For sufficiently small
bandwidths $\Delta\omega \ll \Gamma \scat(\wop)/s_g$, this is
dominated by the scattering at $\wop$, which scales linearly with
diameter in the presence of imperfections as shown in \citeasnoun{Hashemi11}.  On the other hand, for a good single-frequency cloak at
$\wop$, there is a regime $\Gamma \scat(\wop)/s_g \ll \Delta\omega \ll
\Gamma$ where the second term dominates, i.e. where material
dispersion is the limiting factor.  The $C \Delta\omega/\Gamma$ term must therefore also scale linearly
with the diameter in order for $\langle \scat \rangle_{\Delta \omega}/s_g$ to scale linearly for such $\Delta\omega$,
and hence we can conclude that the bandwidth $\Gamma$ of the cloak
generally \emph{scales inversely with diameter}.

This is best illustrated by a simple example.  Consider the case where
$\scat/s_g$ achieves a minimum $f \ll 1$ at $\wop$ with an approximately Lorentzian lineshape of width $\Gamma$, going to~1 at frequencies far from $\wop$:
\begin{equation}
\frac{\scat(\omega)}{s_g} = 1 - \frac{\Gamma^2} {(\omega-\wop)^2 + \Gamma^2}(1-f).
\end{equation}
The integral of \eqref{meanscat} for this $\scat(\omega)$ can be evaluated analytically to obtain
$$
\langle \scat \rangle_{\Delta \omega} = 1 - \frac{1-f}{1+\Delta\omega/\Gamma} \approx f + (1-f) \Delta\omega/\Gamma + O[(\Delta\omega/\Gamma)^2].
$$ Since this must scale linearly with diameter for any $f \ll 1$ and
any $\Delta\omega\ll\Gamma$, it follows that both $f$ and $1/\Gamma$
scale linearly with diameter until cloaking breaks down.

\section{Numerical example}
\label{sec:num}

To illustrate and validate our predictions of this scaling of cloaking
bandwidth, we performed explicit numerical calculations of the scaling for an example bandwidth-limited spherical-cloaking problem (near-perfect at one frequency, but degraded by causal dispersion at other frequencies) in vacuum (with nondimensionalized units $\varepsilon_a = \varepsilon_0 = 1$ and $\mu_a = \mu_0 = 1$).

At an operating frequency $\wop$, we use an exact ``Pendry'' cloak~\cite{Pendry06}: a sphere of radius $R_1$ is surrounded by a
cloak of radius $R_2 > R_1$ that is linearly mapped to an empty sphere
($R_1' = 0$), resulting in materials:
$$\varepsilon_r(\wop)= \mu_r(\wop) = \frac{R_2-R_1}{R_2}\left(\frac{r-R_1}{r}\right)^2$$
$$\varepsilon_\theta(\wop) = \varepsilon_\phi(\wop) = \mu_\theta(\wop) =
\mu_\phi(\wop) = \frac{R_2}{R_2-R_1}.$$  We fixed $R_2 = 1.5 R_1$, so that the cloaking material
parameters are the same for all $R_1$, merely rescaled in space as the object becomes larger or smaller.

By construction, at $\wop$ we have a \emph{perfect} cloak, and the only limiting factor is the bandwidth: we use idealized lossless materials but with
causal dispersion relations that satisfy the Kramers--Kronig
constraints.  In particular, we use a combination of two limiting
cases: a plasma model (a limit of a Drude model as losses go to zero)
and a limit of lossless Lorentzian resonance (corresponding to a
polarization field described by a lossless harmonic oscillator) at a
frequency $\omega_0$.  Combined with the prescribed Pendry values at
$\wop$ from above, this results in dispersion relations:
$$\varepsilon_r(r, \omega) = \mu_r(r, \omega) = 1 - \left[1 - \frac{R_2}{R_2-R_1}\left(\frac{r-R_1}{r}\right)^2\right]\frac{\wop^2}{\omega^2} ,$$
$$\varepsilon_\theta=\varepsilon_\phi=
\mu_\theta=\mu_\phi= 1 +
\frac{R_1}{R_2-R_1}\frac{\omega_0^2 - \wop^2}{\omega_0^2-\omega^2} .$$
The Lorentzian resonance frequency $\omega_0$ can be chosen
arbitrarily; we used $\omega_0 = 2\wop$.

\begin{figure}[t]
\centering
\includegraphics[width=1.0\columnwidth]{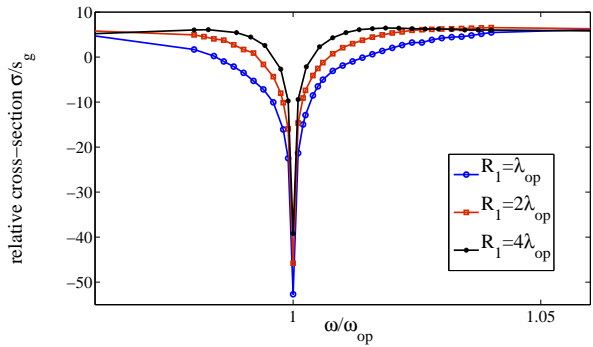}
\caption{Relative cross-section versus frequency for a spherical cloak
  designed to be a perfect Pendry cloak at $\wop$ and showing the
  effects of material dispersion at other frequencies, computed by a
  spectral scattering-matrix method.  As predicted, the cloaking bandwidth
decreases linearly with the object radius, for three object radii relative to $\lop = 2\pi c/\wop$.}
\label{fig:cs-w}
\end{figure}

This geometry was then simulated using a spectral (spherical-harmonic
expansion) scattering-matrix method as described in
\citeasnoun{QiuHu09}.  The continuously varying anisotropic material
parameters are approximated by a large number of piecewise-homogeneous
isotropic layers~\cite{QiuHu09}.  The total scattering cross-section
was computed over a range of frequencies for $R_1 = \lop, 2\lop,
4\lop$ (where $\lop = 2\pi c /\wop$), and is plotted in \figref{cs-w}.
The results in \figref{cs-w} are converged with resolution (the number
of spherical layers) to within a few percent accuracy. At $\wop$, the
cloak should theoretically be perfect, but we obtain a small nonzero
$\sigma/s_g$ ($< 10^{-3}$) due to the discretization errors, which
vanishes with increasing resolution.

As expected, the material dispersion prevents this from being a good
cloak except at frequencies in a narrow bandwidth around $\wop$, and
this bandwidth becomes narrower as the diameter increases.
Quantitatively, if we look at the bandwidth at a fixed $\sigma/s_g$ of
about 1/4 its maximum, we find that the bandwidths for $R_1=2\lop$ and
$R_1=4\lop$ are $\approx 1/2.1$ and $1/4.1$ times the bandwidth for $R_1 =
\lop$, respectively, almost exactly the predicted linear scaling.

[Far from $\wop$, one expects $\scat$ to approach twice the total geometric cross section, in the limit of a large scatterer, because in this limit all of the incident light is scattered.  The factor of two comes from the definition of scattering cross section~\cite{Jackson98}, in which the scattered power appears twice: once as the scattered waves propagating in other directions, and once as the ``shadow'' canceling the forward-propagating wave.  Here, the total geometric cross-section is that of the cloak, $\pi R_2^2$, so we expect $\scat/s_g \approx 10 \log \left[2 (\pi R_2^2) / (\pi R_1^2)\right] \approx 6.5$ (dB) away from $\wop$ for $R_1\to\infty$, and this is roughly what is seen in \figref{cs-w}.]

\section{Concluding remarks}
$\langle \sigma_{tot}\rangle_{\Delta\omega}$
In this work, we extended on our previous paper~\cite{Hashemi11} to study the bandwidth limitation of isolated-object cloaking which is a key limiting factor for this type of cloaking~\cite{Pendry06, Miller06}. Although it was known that perfect cloaking was impossible over a nonzero bandwidth, this result did not seem to exclude the possibility of imperfect cloaking over a finite bandwidth.  Indeed, imperfect finite-bandwidth cloaking is possible, but we have now shown that it is subject to a severe practical constraint: the bandwidth inevitably narrows proportional to the object diameter, given fixed materials.  Although we cannot infer any hard upper bounds on the size or bandwidth of such cloaking without further information about the attainable materials, this result indicates a fundamental challenge in scaling up small experimental demonstrations to larger cloaks.

The use of gain has been proposed to compensate for loss problems in
cloaking~\cite{Han11, Wang10}.  Although gain is necessarily nonlinear and can be detected by a sufficiently strong incident field, in the idealization
of linear gain then many of the techniques in this paper are
complicated by the fact that a linear-gain resonance (the
complex-conjugate of an absorption resonance) would be non-analytic
(and have negative imaginary parts) in the upper-half
complex-frequency plane.  (The average $\scat$ must also be replace by
the average $|\scat|$ or similar, since gain can produce a $\scat <
0$.) However, the bandwidth limitations described in this paper arise
even for idealized materials with negligible dissipation loss, due to
dispersion in the real part of the $\epsilon$ and $\mu$ alone, in
which case even idealized linear gain is inapplicable.  (Furthermore,
as pointed out in our previous work~\cite{Hashemi11}, gain
compensation of absorption must become increasingly perfect as the
object diameter increases, nor does it compensate for scattering from
fabrication disorder.)

Ground-plane cloaking does not suffer from any intrinsic limitation on its bandwidth from causality, nor does our proof apply in that case.  The reason our proof does not apply (at least, in the present form) to ground-plane cloaking is the figure of merit is no longer $\scat$: a ground-plane is actually \emph{designed} to reflect waves, albeit to reflect them in a way that mimics the ground plane.  On the other hand, our previous work~\cite{Hashemi10,Hashemi11} showed that even ground-plane cloaks are increasingly sensitive to imperfections as the size of the object increases. This suggests that ground-plane cloaks should also be increasingly sensitive to material dispersion, for cloaking over a finite bandwidth, as the object size increases, and a challenge for future work is to quantify (or disprove) this relationship.

An alternative direction for cloaking theory (and experiment) is to consider relaxations of the cloaking
problem that might prove more practical. In particular, it would be
valuable to make precise the intuition that the cloaking problem
becomes easier if the incident waves are restricted (e.g. to
plane waves from a certain range of angles) and/or the observer is
limited (e.g. only scattered waves at certain angles are visible, or
only amplitude but not phase can be detected, or sufficiently small time delays are undetectable), since this is arguably
the situation in most experiments. For example, current ``stealth''
aircraft are designed in the radar regime mainly to reduce
back-scattering only~\cite{Nicolai10}. So, it is clear that a sufficiently relaxed cloaking problem is practical even for large objects, and one interesting goal is to find the ``weakest'' relaxation that remains practical at useful scales.

\begin{acknowledgments}
This work was supported in part by the Army Re- search Office through the Institute for Soldier Nanotechnologies (ISN) under contract W911NF-07-D-0004, and by the AFOSR Multidisciplinary Research Program of the University Research Initiative (MURI) for Complex and Robust On-chip Nanophotonics, Grant No. FA9550- 09-1-0704.
\end{acknowledgments}

\section*{Appendix}
Here, we review a known consequence of causality that is the key to
our analysis above: for a passive medium, $\Im \varepsilon (\omega) > 0$ in
the upper half plane $\Im \omega>0$ as proved in~\citeasnoun{Landau84:electro}. A condensed proof of this fact is as follows:
$\Im\varepsilon$ is analytic in $\omega$ in the upper-half plane by causality, and is therefore a harmonic function in the upper-half plane, and so can obtains its minimum
only on the boundary of its domain, except in the trivial case of vacuum where 
it is a constant function ($\Im \varepsilon = 0$). In particular, consider the upper-right quadrant of the complex-$\omega$ plane.  Along the positive real axis, $\Im\varepsilon \geq 0$ for a passive material (in the absence of gain), even for idealized lossless materials. Along the positive imaginary axis, $\Im \varepsilon
 = 0$ since $\varepsilon(-\omega) = \varepsilon(\omega)^*$ for real-valued physical fields~\cite{Jackson98}. As $|\omega|\to\infty$ one must have $\Im\omega \to 0$.  Hence the minimum of $\Im \varepsilon$ along the boundary of the upper-right quadrant is zero and it is strictly positive in the interior.

For physical materials, $\Im \omega > 0$ along the positive real-$\omega$ axis except at $\omega=0$ in order to satisfy the second law of thermodynamics, and this is the usual case in which the above statement is proved~\cite{Landau84:electro}.  However, it is also interesting to consider the
idealized limit of lossless materials (such as a plasma model or a
lossless resonance), in order to study bandwidth-averaged cloaking with idealized materials.  Our proof, above, works even in this case as long as one is a little cautious about the case of poles in $\varepsilon$ lying exactly on the real-$\omega$ axis, in order to exclude the case where $\Im \varepsilon$ diverges to $-\infty$ as the real axis is approached from above.  In particular, we must restrict ourselves to idealized materials that are the limit of physical lossy materials as the losses go zero, so that we are taking the limit as poles approach the real axis from below.  In this case, $\Im \varepsilon$ in the upper-right quadrants is the limit of a strictly positive quantity and hence cannot go to $-\infty$.



\end{document}